\documentclass[]{spie}  
\usepackage[utf8]{inputenc}

\usepackage{amsmath,amsfonts,amssymb}
\usepackage{graphicx}
\usepackage[colorlinks=true, allcolors=blue]{hyperref}

\title{Modeling of the refractive index profile of a femtosecond written waveguide in LiNbO$_3$}

\author[a, b]{Damián A. Presti}
\author[a, c]{Valentín Guarepi}
\author[a, c]{Fabián Videla}
\author[a, b]{Gustavo A. Torchia}
\affil[a]{Centro de Investigaciones Ópticas CONICET – CIC BA - UNLP, Camino Centenario y 506 – M.B. Gonnet, Bs. As., Argentina}
\affil[b]{Depto. de Ciencia y Tecnología, Universidad Nacional de Quilmes, Roque Sáenz Peña 352 – Bernal, Bs. As., Argentina}
\affil[c]{Facultad de Ingeniería, Universidad Nacional de La Plata, Calle 1 y 47 – La Plata, Bs. As., Argentina}

\authorinfo{Further author information: (Send correspondence to Damián A. Presti)\\Damián A. Presti: E-mail: damianp@ciop.unlp.edu.ar}

\begin{document} 
\maketitle

\begin{abstract}

Femtosecond laser pulse systems allows to modify in a precise and permanent way the optical properties of a transparent materials. This process enables the direct writing of guiding structures in materials, commonly known as waveguides, which are the base for optical circuit fabrication.
 
It is our interest to study the main characteristics of the waveguides manufactured by the laser micro-machining technique. Here, an analysis of the resulting refractive index profile has been carried out. This characteristic is essential for the design and simulation of integrated optical circuits. In particular we have developed our research on the study of light coupling in a pair of type II waveguides made in Lithium Niobate (LiNbO$_3$).These experimental backgrounds provide us with elements to adjust and test the retrieved profile. Taking into account different distance between tracks and writing energies, it is well known that the coupling length changes and the coupling ratio too.
Then this study allows us to reconstruct the refractive index profile according to its manufacturing conditions. Modeling of the refractive index distribution profile is a key parameter to perform beam propagation mode simulations (BPM) to achieve more realistic results. So, by means of this method it is possible to obtain a general procedure to describe the characteristics of these kinds of waveguides. As a model test, integrated waveguides were built to corroborate their light coupling. In a first stage it is designed through BPM simulations then it is manufactured in an X-cut LiNbO$_3$ crystal in order to check its operation according to the simulations carried out. 
\end{abstract}


\section{INTRODUCTION}
\label{sec:intro}  

Integrated optics has become one of the most relevant topics in the field of photonics and semiconductor device research. The use of fiber optic technology, the growing potential of optical communication systems and the new sensors applications in multiple areas, have emphasized the need for integrated optical components (OI), such as couplers, modulators, switches, filters, detectors, sensors, etc., that are reliable and accurate. In addition, the technological capability to manufacture multiple waveguides shapes with relative differences in the refractive index and the use of semiconductor optical materials has introduced a variety of highly compact components that are suitable for optical integration. 

The design of optimized integrated optical components requires a detailed understanding of the various characteristics of electromagnetic propagation in these structures, among other factors 
that affect the characteristics and define the functionality of the devices. The method of analysis that can provide a complete solution to the characteristics of a component will improve the optimization of system performance.
In addition, a precise method to emulate the operation of components and circuits allows ways to explore new ideas for devices and systems without the cost of manufacturing and testing. This demand for methods and models of analysis, which are general and versatile, made that in the early 90s began to discuss and develop the first simulation programs \cite{Chu1990}. Currently CAD (Computer Aided Design) tools and simulators play a fundamental role in the current advances and previous achievements that have been made, in the photonic area, more specifically in integrated optics.

An optical waveguide in its transverse direction can generally span between a few wavelengths until it is below it and the transverse profile can be classified according to the distribution of the index in the guide region \cite{Okamoto2006}. These parameters are strongly linked to manufacturing technology and material. For a stepped index waveguide, the guide region has a uniform refractive index distribution, while the refractive index for a gradual index waveguide typically decreases gradually from the center of the guide region to the external region. These differences will be directly associated with the waveguide manufacturing process. 

Currently there are multiple commercial programs dedicated to the simulation of optical circuits. They basically perform an optical analysis on a specific geometry. First, for a given uniform structure, they determine their normal mode distribution and their associated propagation constant. Second, they determine the behavior of the optical signal propagating in the given geometry. The solution to these problems is found by solving Maxwell's equations in the domain subject to the boundary conditions dictated by the geometries. For uniform structures, a large number of methods have been developed and implemented. The effective index method, the variational method, the mode-matching method, finite-element method and finite-difference method are some of the examples. For non-uniform structures, the commonly used approaches are: Coupled-Mode Theory (CMT), the Beam Propagation Method (BPM) and the Finite Difference Time Domain (FDTD) method. The choice of the most efficient method of resolution will depend on the waveguides of our integrated optical circuit and its complexity in terms of geometry and index profile \cite{Chu1990, Parfenov2016, Vitkovskiy2008}.

The main objective of this work is to present the use of these simulation techniques. We will focus specifically on the study of waveguides manufactured by direct laser writing process in Lithium Niobate (LiNbO$_3$) \cite{Toney2015, Bazzan2015, Courjal2018}. In particular, this technique has been developed as an alternative method of manufacturing optical circuits in optical materials with great potential. This method is based on the interaction of ultrashort pulses of laser light and optical materials. By a suitable choice of laser irradiation conditions, it is possible to achieve an increase in the refractive index highly localized in said volume \cite{hirao1996, chen2013, Osellame2012, lv2016}. In this way, the optical properties of a small volume inside a transparent material can be modified in a precise and permanent way. This feature allows the direct writing of an optical waveguide. This waveguide has the peculiarity of having a non-uniform refractive index profile and dependent on the writing parameters. It is of our interest to study the characteristics of the waveguides manufactured by the laser micro-machining technique. For this, an analysis of the resulting refractive index profile will be carried out. This feature is of great relevance for the design and simulation of integrated optical circuits. 

We will make an approximation (by mean of simulations) to the refractive index profile, and then adjust it experimentally based on the study of light coupling in type II waveguides manufactured in Lithium Niobate (LiNbO$_3$) \cite{szameit2007, Neyra2014}. Considering different distances between tracks (double track guides) and taking into account the variation of this writing parameter, it is possible to find a modification of the refractive index profile in each case, which will lead to a change in the coupling length. This study allows the reconstruction of the refractive index profile according to its manufacturing conditions. From the analysis of experimental data in each of the manufactured guides, it is expected to develop a mathematical model to describe the characteristics of waveguides according to the parameters used in the manufacturing process. For the process of reconstruction and verification of the model, we will rely on the waveguides simulation using the BPM wave propagation calculation method in the Rsoft commercial optical simulation software.

\section{MATERIAL AND METHODS}

\subsection{Fabrication: Writing process}
\label{sec:escritura}

A Titanium Sapphire laser system was used to manufacture the Type II guide set. It emits pulses of 150 femtoseconds, centered at a wavelength of 800 nm and with a repetition frequency of 1 KHz. A value of 0.7 $\mu J$ of laser energy was used at a rate of 50 $\mu m/sec$ to define the waveguides. The material to record them was Lithium Niobate crystal (LiNbO$_3$) cut x. These values were pre-set to guarantee an adequate writing fluency and obtain Type II guides within the material \cite{Peyton2018}. 

Taking into account the coupling study that we carried out, the waveguides should be single and double track \cite{chen2013}. A set of single track and 6 double track waveguides were recorded maintaining the same writing characteristics and only varying their distance between tracks (gap) for the second ones. The recorded gap values were: 25, 27, 29, 31, 33 and 35 $\mu m $. Varying only one parameter (gap), we isolate the study of light coupling over all the possible structural changes of the manufacturing process.

\subsection{Refractive index profile (Stage 1): Approach to the model}

\begin{figure} [ht]
   \begin{center}
   \begin{tabular}{c} 
   \includegraphics[height=3cm]{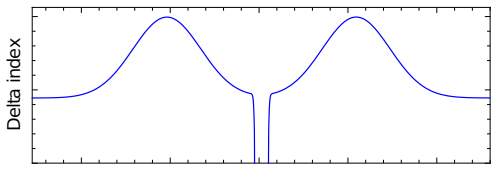}
   \end{tabular}
   \end{center}
   \caption[example] 
   { \label{deltaindex_model} 
Linear delta distribution of the refractive index according to the Equation \ref{indexexpr}}
   \end{figure}

As already mentioned, writing by femtosecond laser pulses generates a waveguide with a non-uniform refractive index profile. Several previous works have focused on this phenomenon \cite{Tejerina2013, Biasetti2013}. A first approximation of the refractive index profile of this type of waveguides can be represented as a sum of Gaussian functions. We start  from a first general approximation, and then adjusted the model according to the experimental data obtained. In this first stage we use waveguides manufactured as simple track. For this, we take the sum of 3 Gaussian functions for each step of laser writing: one negative where the track is generated (decrease of the refractive index) and two positive ones for the external lobes where the guiding effect occur because of the increase of the refractive index. The function that describes it is the Equation \ref{indexexpr} and is schematized in the Figure \ref{deltaindex_model}.

     \begin{equation}
       f(x,y)=e^{-c_1(d+e.x)^2} +a e^{-c_2 y^2}e^{-c_1 y^2}e^{-c_1(-d+e.x)^2} -a_t e^{-b_t x^2}e^{-c_2 y^2}
        \label{indexexpr}
     \end{equation}

\begin{figure} [ht]
   \begin{center}
   \begin{tabular}{c} 
   \includegraphics[height=5cm]{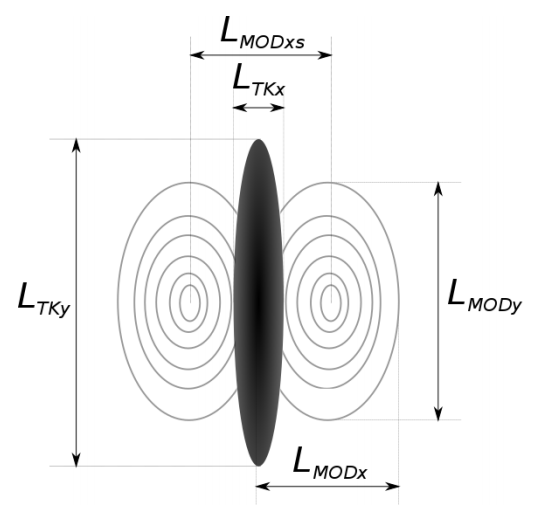}
   \end{tabular}
   \end{center}
   \caption[example] 
   { \label{index_model} 
Spatial distribution scheme of the refractive index for a single track waveguide manufactured by direct laser writing (front view). With $L_{TKx}, L_{TKy}, L_{MODx}, L_{MODy}$ and $L_{MODxs}$ as adjustment coefficients}
   \end{figure}

Where $a, a_t, b_t, w, c_1, c_2$ and $d$ are adjustment coefficients for the model. In Figure \ref{index_model} we can see other representation of  the equation \label{indexexpr} as the contour map, where two dimensional details can be appreciated. Featuring lengths have been indicated: the length in $X$ of the track ($L_{TKx}$), the length in $Y$ of track ($L_{TKy}$), the length in $X$ of mode ($L_{MODx}$), the length in $Y$ of mode ($L_{MODy}$) and the length between modes ($L_{MODxs}$) 
These lengths can be adjusted by mean of
the coefficients associated with Equation \ref{indexexpr}. It is clear that obtaining these parameters will bring us closer to the adjusted  model. 

Regarding the measurements, the track lengths is the first, since it is enough to make it  a calibrated microscope on the edge of the glass. The track, having a highly negative refractive index can be observed and measured without great complexity. For the calculation of the featuring lengths, we analyze the shape and intensity of the output modes. We perform this measurement by injecting light through one of its edges and aligning a beam profile analyzer to the output. To couple the light inside the structure, a laser diode was used, while at the output a microscope objective was used to collect the near field light from the output. Taking as a reference the measurements already obtained from the track, we can make a first approximation of the values that involve the modes. This is a first approximation between the model and experimental result, comparing the output mode measured against the simulated one, whose similarity is evaluated inspecting its shapes. The closer is the similarity between the experimental and the model, the better will converge in Stage 2 of adjustment.

\subsection{Refractive index profile (Stage 2): Model adjustment by coupled modes}

When two straight waveguides are close (ie, separated by a tens of micrometers), the propagation modes can generate optical coupling. This is due to the superposition that occurs between the evanescent fields of one waveguide with the other. The parameter that determines the coupling between the waveguides is called the coupling constant $k$. This parameter is determined by the superposition between the guided fields of the waveguides and their refractive indexes, and decreases as the separation between the waveguides increases. This is because the fields in the opposite direction to the waveguides tend to zero at infinity. If the two waveguides have the same characteristics, the coupling efficiency between them has a maximum value when the power of one is transferred completely to the other. This energy transfer occurs periodically and is characterized by the coupling length $L_c$, which is the shortest length at which the power of one waveguide is completely transferred to the other waveguide. For a coupling efficiency of 100 \%, the coupling constant $k$ and the coupling length $L_c$ are related by $L_c = \pi / 2k$ \cite{Okamoto2006}. This coupling characteristic strongly depends on the refractive index profile of the guides. A good model of this should not only reflect the output modes of a waveguide, but also reflect the same $L_c$. This second stage consists of the adjustment and validation of the parameters established by Stage 1 using the coupling parameters.

\begin{figure} [ht]
   \begin{center}
   \begin{tabular}{c} 
   \includegraphics[height=5cm]{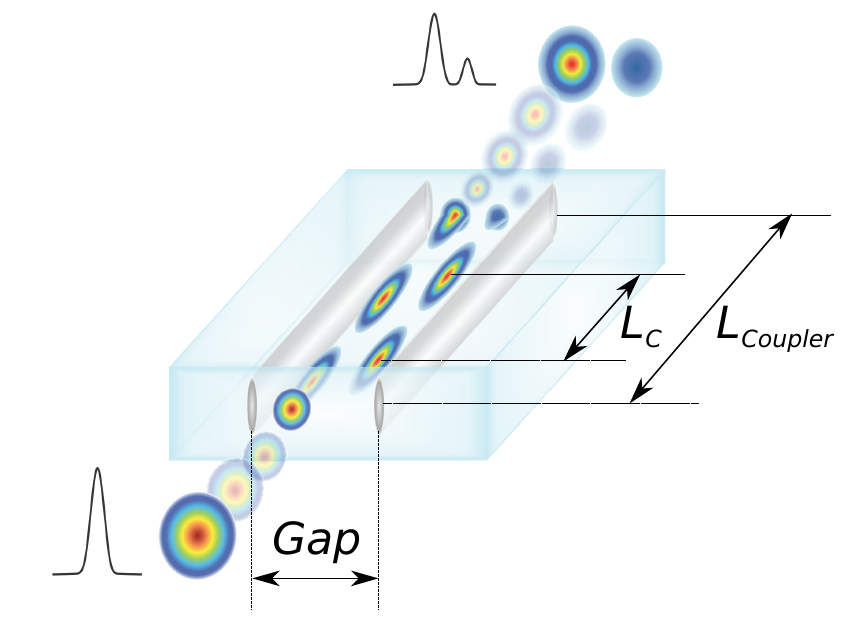}
   \end{tabular}
   \end{center}
   \caption[example] 
   { \label{acople_esquema} 
Schematic representation of a pair of waveguides fabricated by direct writing process with a femtosecond laser inside a LiNbO$_3$ crystal and the coupling modes between the two waveguides}
   \end{figure} 

In Figure \ref{acople_esquema} it is possible to observe a scheme of double track Type II waveguides where light is coupled by one of its ends. The output varies depending on the total coupling length ($L_{coupler}$) due to the effect of coupled modes. In turn, as we mentioned, $L_C$ depends on the length of the gap. Therefore we can say that the proportion of light in the output modes will vary depending on the Gap length. For all the waveguides of the recorded set, it is expected to obtain different combination of output intensities. If the model predict them  then, can be considered reliable. 

For the simulations between two Type II straight waveguides, the BPM method was used, where the refractive index profile was loaded according to the Equation \ref{indexexpr}. An iterative code was programmed to perform the simulations of our experimental system using the desired gaps and altering the equation coefficients of the model. When the solutions converge with the experimental measures, we obtain the final results of our adjustment variables, and thus our adjusted model.

\section{RESULTS AND DISCUSSION}

During the experimental process, light coupling tests were carried out for each of the recorded waveguides. For this, a 980 nm laser diode coupled to optical fiber was used at the input, while at the output a 10x microscope objective aligned with a CCD camera was used to characterize the shape and intensity in proportion to the output modes. This means, to obtain the output distribution intensity whose values will be expressed in a relative form. To obtain the refractive index profile we will detail the results according to the stages mentioned above.

\subsection{Results of Stage 1: Approach to the model}

During this stage, the coupled waveguides were the simple track. The main objective is to obtain the guided mode to the output experimentally, to then approximate the refractive index model to this. 

\begin{figure} [ht]
   \begin{center}
   \begin{tabular}{c} 
   \includegraphics[height=5cm]{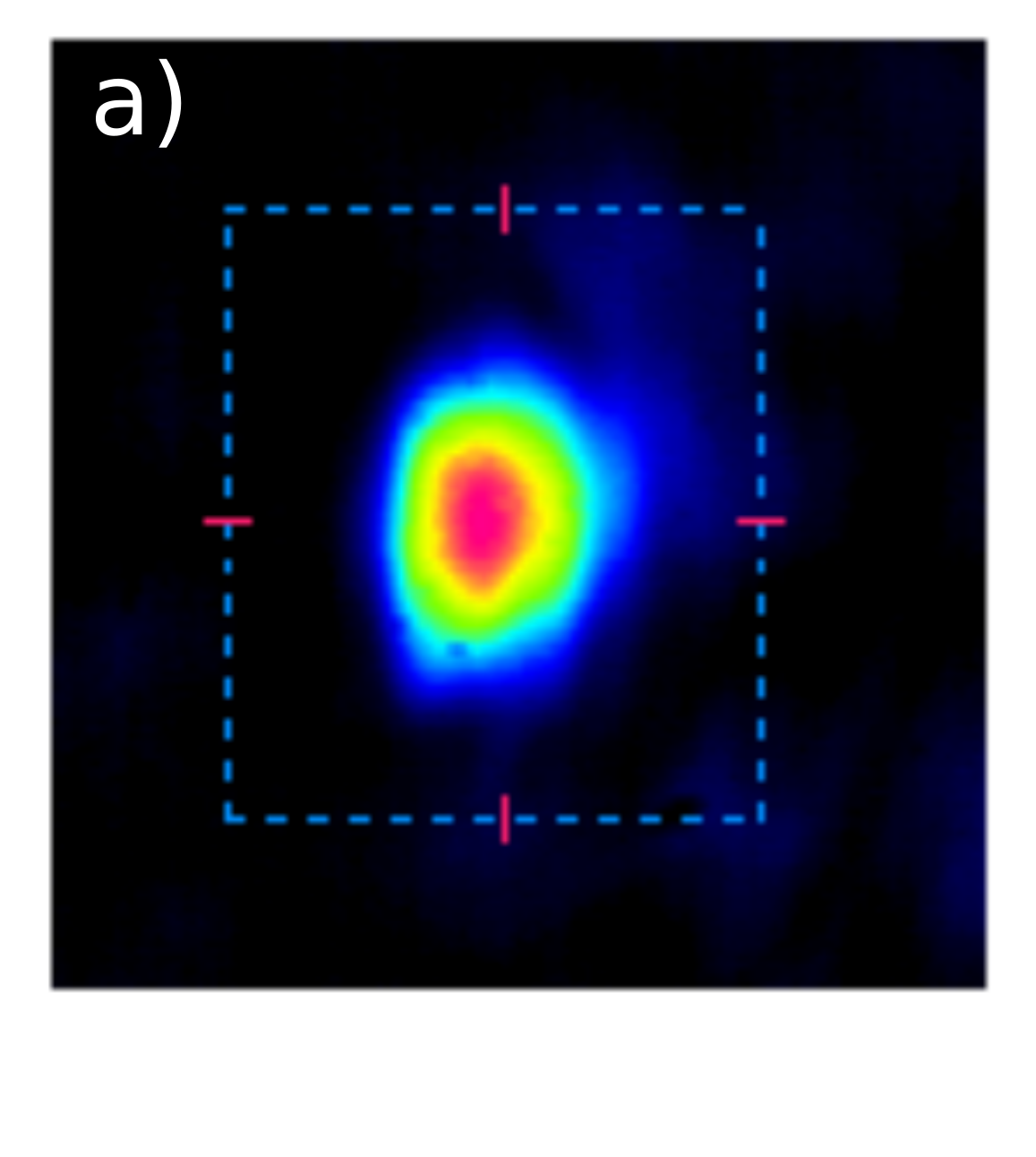}
   \includegraphics[height=5cm]{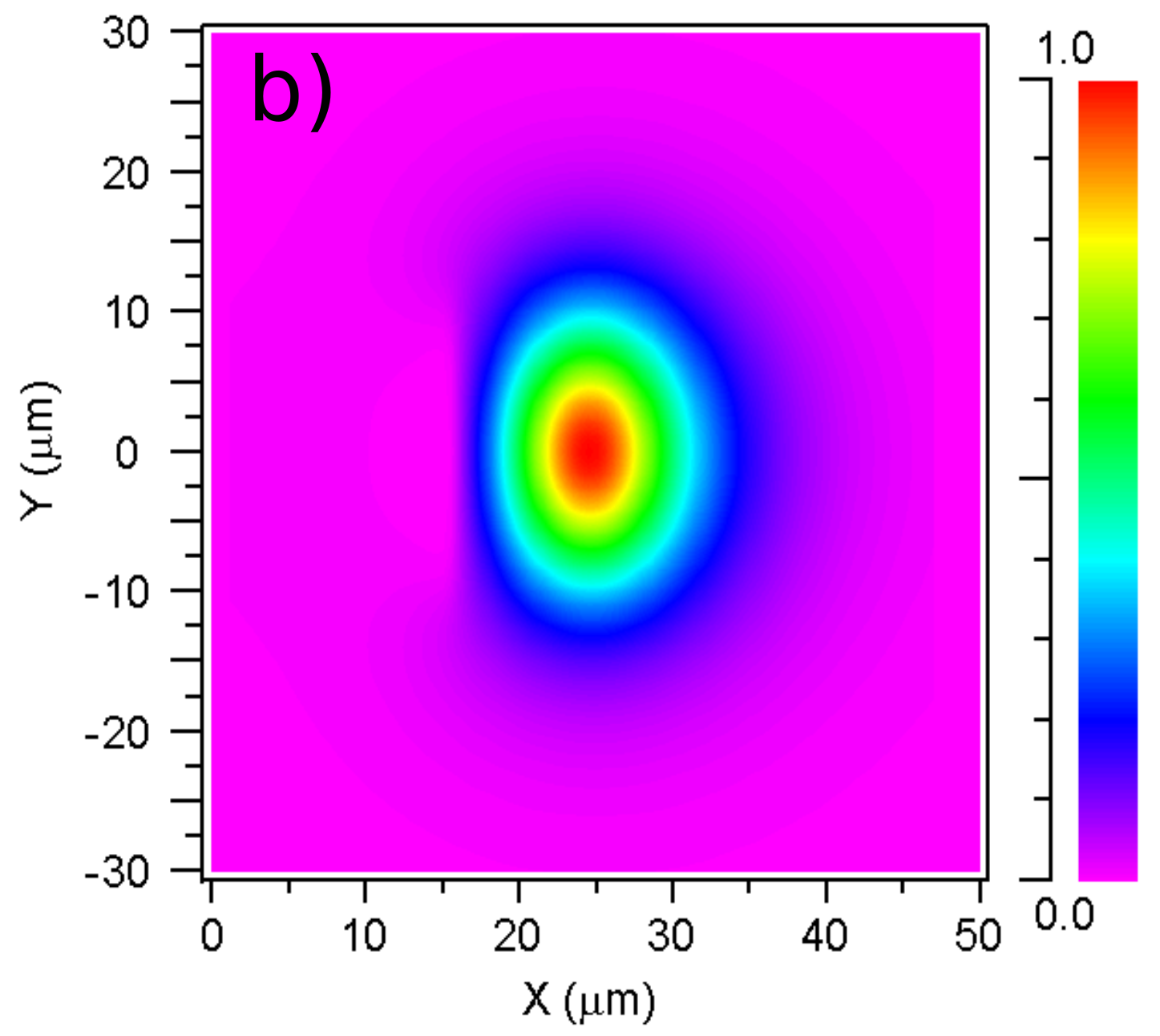}
   \end{tabular}
   \end{center}
   \caption[example] 
   { \label{modos_S_track} 
Coupled mode for a simple track waveguide: a) Experimental, b) Simulation}
\end{figure}

In Figure \ref{modos_S_track} a) we can see the mode experimentally obtained by the CCD camera. In order to observe the track but with the same setup,it is necessary to saturate the image, because the track is unable to propagate light, so, the contrast is increased. Thus, we can get the measurements of  $L_{TKx}$ and $L_{TKy}$ of the track. On the other hand we can also obtain its measurements through a calibrated microscope. Therefore, by correlating these measurements with the capture of the mode and the track, we can scale the actual size of the output mode. These values are the ones that we will replicate in a computational way. 

As for the simulation process, the CAD of the coupled guide circuit was performed, the proposed refractive index profile (Ec. \ref{indexexpr}) was loaded and the values of its coefficients were approximated to the dimensions of the mode obtained. The figure \ref{modos_S_track} b) shows the simulation obtained from the mode at the output of the waveguide.

\subsection{Results of Stage 2: Model adjustment by coupled modes}

This second stage iteratively adjusts the model proposed in Stage 1 until converging to the experimental coupling results. In this case, the tests were carried out with the double track waveguides. With a Gap large enough, the modes are coupled and it is possible to achieve measures of intensities in both guides separately. The experimental coupling measurements  values of intensities are those shown in Table \ref{Tabla_acoples}.

\begin{table}[ht]
\caption{Experimental relative intensity measurements (\%) for each pair of waveguides of the manufactured set} 
\label{Tabla_acoples}
\begin{center}       
\begin{tabular}{|c|c|c|} 
\hline
\rule[-1ex]{0pt}{3.5ex}  Gap & Waveguide \#1 & Waveguide \#2 \\
\rule[-1ex]{0pt}{3.5ex}  [$\mu m$] & intensity [\%] & intensity [\%] \\
\hline
\rule[-1ex]{0pt}{3.5ex}  25 & 92,7 & 7,3  \\
\hline
\rule[-1ex]{0pt}{3.5ex}  27 & 10,8 & 89,2  \\
\hline
\rule[-1ex]{0pt}{3.5ex}  29 & 80,2 & 19,8  \\
\hline
\rule[-1ex]{0pt}{3.5ex}  31 & 43,4 & 56,6  \\
\hline 
\rule[-1ex]{0pt}{3.5ex}  33 & 18,9 & 81,1  \\
\hline
\rule[-1ex]{0pt}{3.5ex}  35 & 40,8 & 59,2  \\
\hline
\end{tabular}
\end{center}
\end{table}

\begin{figure} [ht]
   \begin{center}
   \begin{tabular}{c} 
   \includegraphics[height=5cm]{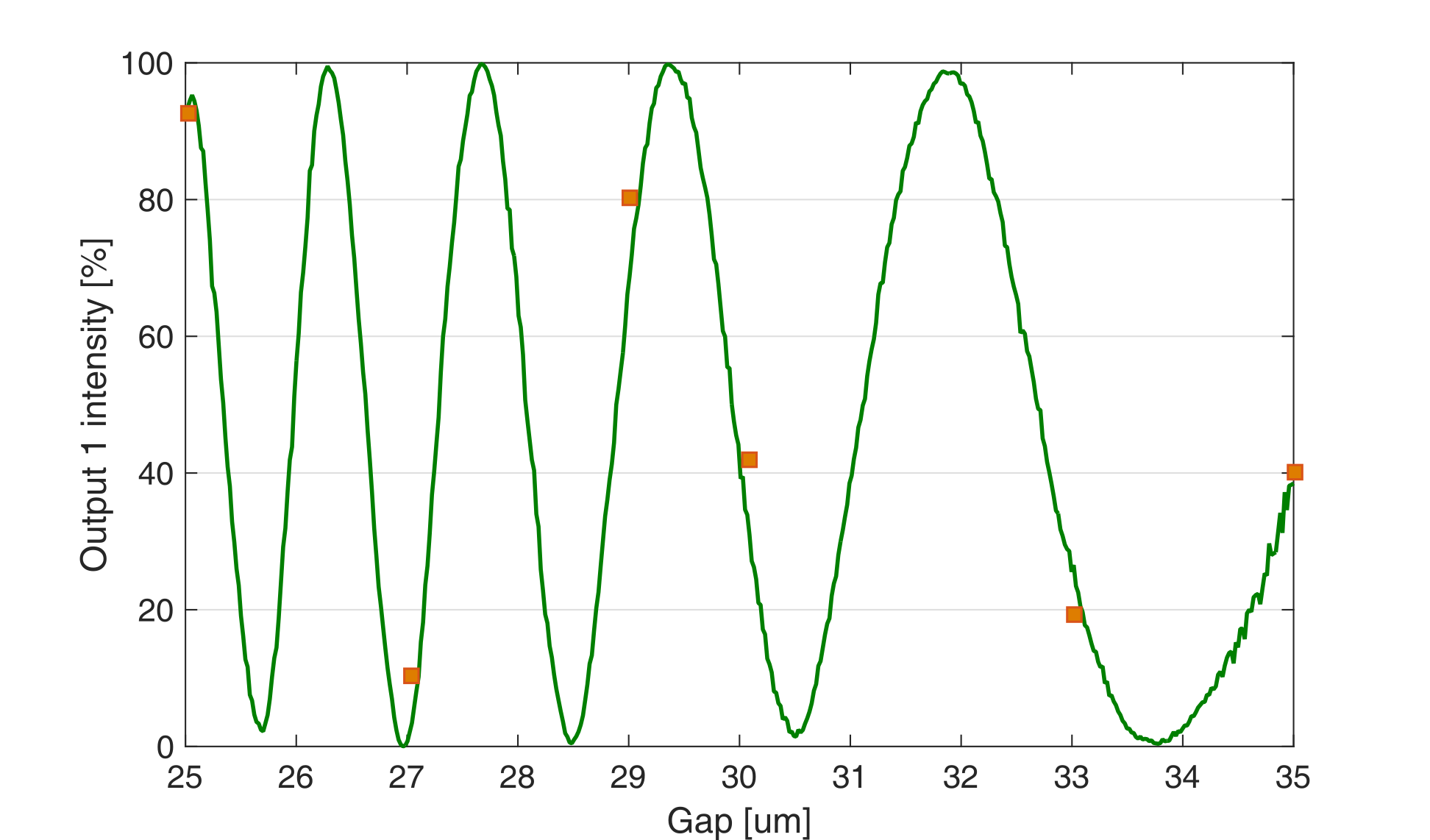}
   \end{tabular}
   \end{center}
   \caption[example] 
   { \label{acople_curva} 
Relative output power intensity in waveguide 1 as a function of Gap length. In orange cubes experimental measurements, in green results of the refractive index profile model by simulation}
   \end{figure}

To obtain the adjustment coefficients, we again use the simulated model in the iterative process. Initially we set the coefficients at values previously obtained  by the Equation \ref{indexexpr}. This guarantees us to modify the parameters of the refractive index profile, without getting too far away from the first adjust. The iterative process will continue as long as necessary, until the intensities predict by the model (through simulations) converges to the coupling values obtained experimentally. In Figure \ref{acople_curva} we can observe the result of this process taking the coupling percentage of a single waveguide.

The green curve represents the coupling values estimated by simulations in a continuous range of gap values. It is expected that the curve goes from a maximum to a minimum (100 \% of coupling to 0 \%) and as the Gap increases its period as well (since L$_C$ increases). 
In orange cubes are represented the experimental measures. The predicted behaviour shows good agreement with the experimental ones.

Finally, for this model, the obtained coefficients are those expressed in Table \ref{tabla_coeficientes}.

\begin{table}[ht]
\caption{Resulting values for the coefficients of the refractive index profile equation} 
\label{tabla_coeficientes}
\begin{center}       
\begin{tabular}{|c|l|} 
    \hline  
    Coefficients & Values\\
    \hline
    $a$ & 0.045\\
    \hline
    $a_t$  &  60 \\
    \hline
    $b_t$ & 20 $\mu$m$^{-2}$ \\
    \hline
    $c_1$ & 0.012 $\mu$m$^{-2}$ \\
    \hline
    $c_2$ & 0.0125 $\mu$m$^{-2}$\\
    \hline
    $d$ & 18 $\mu$m\\
    \hline      
\end{tabular}
\end{center}
\end{table} 

\begin{figure} [ht]
   \begin{center}
   \begin{tabular}{c} 
   \includegraphics[height=8cm]{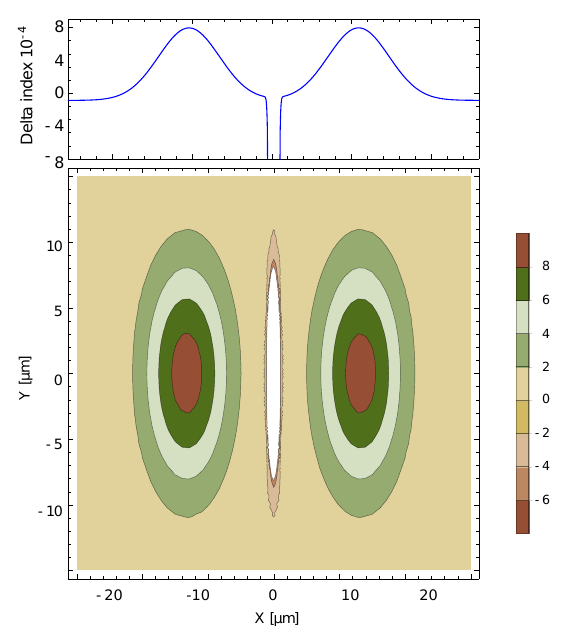}
   \end{tabular}
   \end{center}
   \caption[example] 
   { \label{perfil_final} 
Refraction index profile model for a simple track Type II waveguide by laser direct writing. Linear and spatial distribution of refractive index}
   \end{figure}

Returning to the equation \ref{indexexpr} and replacing the obtained coefficients, we obtain the final refractive index profile. In Figure \ref{perfil_final} we can observe its linear and spatial distribution.

\section{CONCLUSIONS}

Femtosecond laser writing is currently used as an alternative technique for the fabrication of optical circuits, among other applications. Given the relative advantages of respect to conventional technologies, it is of our interest to study its characteristics and potentialities. In particular, a precise method allows us to simulate the operation of an optical circuit manufactured by this technique. This give rise to more optimized designs, as well as  ways to explore new concepts.

This procedure was presented to obtain, from the analysis of experimental data, a model for a non uniform refractive index profile for Type II waveguides manufactured using the direct femtosecond writing technique. This model describes the characteristics of waveguides considering the parameters used in the manufacturing process. 

The model reconstruction and verification process was based on simulation using the BPM wave propagation calculation method for waveguides manufactured in Lithium Niobate (LiNbO$_3$). 

The simulation model obtained highlights the possibility of improving preliminary designs, as well as the ability to replicate the study for other materials or conditions using the same manufacturing technique.

\acknowledgments 
 
This work was partially supported by the Agencia Nacional de Promoción Científica y Tecnológica (Argentina) under the projects PICT-2016-4086  and PICT-2017-0017. By Universidad Nacional de Quilmes under the project PPROF-901-2018. DAP, VG and GAT are with CONICET, Argentina. FV belongs to the Comisión de Investigaciones Científicas (Buenos Aires, Argentina).

\bibliography{report} 
\bibliographystyle{spiebib} 

\end{document}